# Spin-orbit torque switching of synthetic antiferromagnets

Chong Bi[1†], Hamid Almasi[1], Kyle Price[1], Ty Newhouse-Illige[1], Meng Xu[1], Shane R. Allen[2], Xin Fan[2] and Weigang Wang[1*]

[1]Department of Physics, University of Arizona, Tucson, Arizona 85721, USA

[2]Department of Physics and Astronomy, University of Denver, Colorado 80208, USA

Abstract

    We report that synthetic antiferromagnets (SAFs) can be efficiently switched by spin-orbit torques (SOTs) and the switching scheme does not obey the usual SOT switching rule. We show that both the positive and negative spin Hall angle (SHA)-like switching can be observed in Pt/SAF structures with only positive SHA, depending on the strength of applied in-plane fields. A new switching mechanism directly arising from the asymmetric domain expansion is proposed to explain the anomalous switching behaviors. Contrary to the macrospin-based switching model that the SOT switching direction is determined by the sign of SHA, the new switching mechanism suggests that the SOT switching direction is dominated by the field-modulated domain wall motion and can be reversed even with the same sign of SHA. The new switching mechanism is further confirmed by the domain wall motion measurements. The anomalous switching behaviors provide important insights for understanding SOT switching mechanisms and also offer novel features for applications.

[†]cbi@email.arizona.edu

[*]wgwang@physics.arizona.edu

**I. Introduction**

Electrical manipulation of magnetization is a crucial step for encoding data in spintronic memory and logic devices. It is usually achieved through the spin transfer torque (STT) effect [1] generated by a spin-polarized current in a spin-valve or magnetic tunnel junction (MTJ) structure. In recent years, spin-orbit torques (SOTs), a new type of spin torques driven by in-plane currents flowing in heavy-metals (HMs) [2–6], topological insulators [7–9] or antiferromagnets [10,11], have emerged as a more efficient way to manipulate magnetization. SOTs have been successfully employed to switch magnetization [2–4,8,10,12–18], drive domain wall (DW) motion [19,20] and excite spin-torque nano-oscillators [21,22]. In many applications such as magnetic random access memory (MRAM), SOTs have advantages over STTs due to their higher efficiency and the ability to switch a MTJ without passing a large current through the tunnel barrier. However, as a fundamental question, the underlying SOT switching mechanism is still under debate. Moreover, the contributions of various interfacial effects, such as the Rashba effects, spin Hall effects (SHEs) and Dzyaloshinskii–Moriya interaction (DMI) to the SOT switching also remain elusive. One widely accepted SOT switching mechanism is based on the macrospin model [4], in which the SOT nucleates initial domains through the macrospin model and switches the entire ferromagnet by subsequent domain expansion [4,12,23–25]. In this model, SHEs dominate the SOT switching, which obeys the rule shown in Figs. 1(a) and (b). The final magnetization switches to the direction determined by $\mathbf{H} \times \mathbf{\sigma}$. Here, $\mathbf{H}$ is the effective in-plane field that is provided by an external magnetic field [2–4,8,12–15,18], exchange bias or interlayer coupling [10,17], and $\mathbf{\sigma}$ is the spin polarization injected from adjacent materials. The spin Hall angle (SHA) of adjacent materials determines the direction of $\mathbf{\sigma}$ and thus the switching direction for a given $\mathbf{H}$. Usually the adjacent materials can be classified into two basic types with a positive SHA, such as Pt [4], (Bi, Sb)Te [8], and PtMn [10], and a negative SHA, such as Ta [3] and W [26]. Therefore, $\mathbf{\sigma}$ and the switching directions in Pt/ferromagnet (Pt/FM) and Ta/FM structures are opposite [2–4,8,10,12–17]. So far all reported SOT switchings are the SOT switchings of a single ferromagnet [2–4,12,14,18,23–25,27], in which reversing either current or in-plane field is necessary to switch magnetization [2–4,12,14,18,23–25,27] and the switching rule shown in Figs. 1(a) and (b) is well obeyed.

In high-density MRAMs, synthetic antiferromagnets (SAFs) are widely adopted in MTJs as the reference layer [28–30] and even as the free layer [31,32] to improve the thermal stability of MTJs and reduce the coupling field between the reference and free layers. A SAF with strong interlayer coupling can generally be regarded as a FM layer with the effective magnetization of $\mathbf{M_{eff}} = \mathbf{M_A} + \mathbf{M_B}$, where $\mathbf{M_A}$ and $\mathbf{M_B}$ are the magnetization of two coupled FMs (see Supplemental Material). According to the macrospin model, one will expect that the SOT switching of HM/SAF to be similar to that of HM/FM, since the SHA and corresponding SOTs are exactly the same in the two systems. Up to date, the SOT switching of SAFs and its efficiency have not been investigated.

Here we demonstrate that the SAFs can also be efficiently switched like a single ferromagnet. Surprisingly, the switching scheme of SAFs does not obey the usual switching rule shown in Figs. 1(a) and (b). It is shown that the SOT switching direction of SAFs can be reversed depending on the strength of applied in-plane fields even with the same sign of SHA. These results indicate that the switching of SAFs can be achieved without any direction changes of the applied in-plane field and current, contrary to the switching of a single ferromagnet [2–4,12,14,18,23–25,27] in which the direction of either current or in-plane field has to be reversed. The observed anomalous magnetization switching (AMS) behaviors invalidate the conventional macrospin model and prompt a new understanding of SOT switching. To explain the AMS behaviors, we then propose a new SOT switching mechanism directly arising from the asymmetric domain expansion/contraction due to the field-modulated chiral DW motion. The new switching mechanism suggests that the SOT switching direction is only determined by the in-plane field modulated relative velocity between ↑↓ and ↓↑ domains ($V_{RD}$) [19,20], regardless of the initially nucleated domains through the macrospin model, and thus does not directly depend on the sign of SHA. The current-driven DW motion measurements further confirm this switching mechanism and demonstrate that the AMS arising from the unique chiral DW motion due to the special magnetization configuration of SAFs. The unique DW motion in SAFs causes that the switching directions deviate from the macrospin model, clearly clarifying the new SOT switching mechanism that has not been revealed in the previously reported SOT switching of a single ferromagnet. These results highlight the DMI effects that determine the chiral DW motion during the SOT switching and also provide a guideline for optimizing SOT switching in applications. Furthermore, this novel switching behavior combined with tunable interlayer coupling [33–36] could also enable many new SOT-related applications.

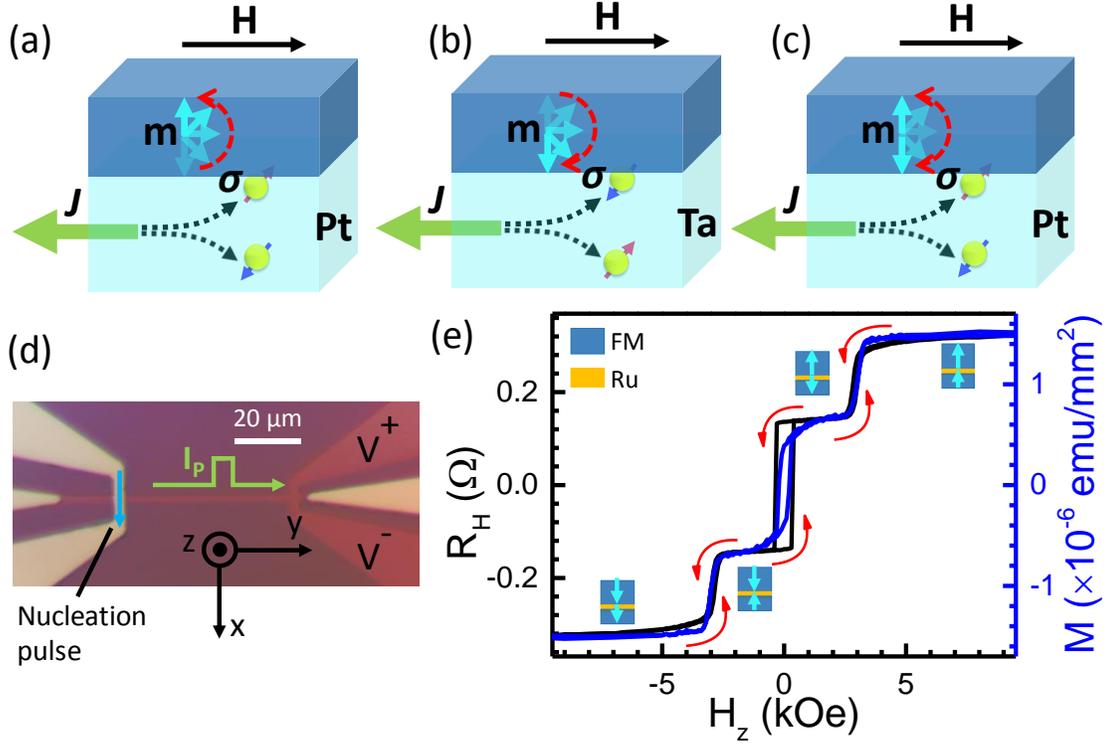

FIG. 1. The SOT switching with a (a) positive or (b) negative SHA. $J$ is the injected current density, and **m** is the unit magnetization vector of adjacent magnets. **m** switches to up or down for the positive or negative SHA, respectively. (c) SOT induced anomalous switching observed in this work, in which **m** can be switched to both up and down states under the SOT with the same sign. (d) Top-view of a Hall bar structure showing the configurations of electrical measurements and coordinate system. (e) Magnetic properties of the Pt/SAF structure characterized by AHE (black) and VSM (blue). The red arrows show the switching sequence of magnetization. The insets show the magnetization configuration at each field stage.

## II. Experimental details

*Sample fabrication*: The samples employed in this work have the structure of Si-wafer/SiO$_2$ (300nm)/Pt (4nm)/BML/Ru $t_{Ru}$/TML/Ru (0.6nm)/SiO$_2$ (10nm), where BML (bottom magnetic layer) is Co (0.6nm), TML (top magnetic layer) is Co (0.4nm)/Pt (1nm)/Co (0.4nm)/Pt (1nm)/Co (0.4nm), and $t_{Ru}$ is the thickness of the Ru spacer layer in the range of 0 - 1.5 nm. Here we adopted a symmetric TML to minimize the SOTs from the inside Pt layers. Control samples with a thicker BML were also fabricated. All the stack structures were deposited on Si/SiO$_2$ (300 nm) substrates by magnetron sputtering. The deposition rates for each layer are: Pt 0.05 nm/s, Co 0.018 nm/s, Ru 0.01 nm/s, and SiO$_2$ 0.074 nm/s. The base vacuum was better than $1.5\times10^{-8}$ torr before sputtering. The samples were then patterned into Hall bar structures with a feature width of 2.5 µm, as shown in Fig. 1(d). To monitor DW motion, an orthogonal DW nucleation line with a width of 3 µm was directly deposited on the top of each Hall bar structure. The distance between the nucleation line and the voltage bars is 50 µm. The magnetic properties of fabricated continuous films were measured by vibrating sample magnetometry (VSM) measurements, and the anomalous Hall effect (AHE) measurements were performed in patterned Hall bar structures. The ferromagnetic (FM) or antiferromagnetic (AFM) interlayer coupling for each sample was determined by the combination of AHE and VSM measurements.

*Electrical measurements*: The current pulses and a 0.3 mA dc current for sensing anomalous Hall resistance ($R_H$) were applied by the same Keithley 6220 current source. The Hall voltage was monitored by a Keithley 2000 multimeter. For all SOT switching related measurements (by sweeping current pulses or external fields), a 1-ms current pulse was applied first. After waiting 3 s, the applied external field was then removed and a 0.3 mA dc sense current was applied to detect magnetization states after each current pulse. For DW motion measurements, the magnetization of the Hall bar was first initialized to a uniform up or down state by a positive or negative 6 kOe perpendicular field, respectively. After that, a 1 ms nucleation current pulse was applied by an independent Keithley 2400 Sourcemeter to create domain nucleation. The amplitude of nucleation pulse was 120 mA. The nucleation current was negative for up initial states and positive for down initial states. To increase the probability of domain nucleation, an assisted perpendicular field, a bit smaller (typically 50 Oe) than reversal switching fields, was applied during the nucleation current pulse. After domain nucleation, a current pulse was then applied by the Keithley 6220 current source to drive DW motion. The $R_H$ was measured after the current pulse to determine if the DW had arrived at the voltage bars. The length of the applied current pulse varied between 1 ms and 10 s. We chose the proper injected current densities to make sure that the time for DW motion between the nucleation line and the voltage bars was within 1 ms to 10 s. The detailed measurement process is given in Supplemental Material. Hereafter, we mainly present the experimental results from the sample with $t_{Ru}$ = 0.66 nm that shows strong AFM coupling.

## III. Results

The schematic process of the SOT induced anomalous switching in this work is illustrated in Fig. 1(c). In contrast to the conventional switching in Pt/FM (Fig. 1(a)) and Ta/FM (Fig. 1(b)), the switching sign in a Pt/SAF structure can reverse even at the same external field direction. Figure 1(e) presents the AHE and VSM results, both of which show three clear perpendicular switching loops, indicating an AFM interlayer coupling as well as a strong perpendicular magnetic anisotropy (PMA) in the sample. The switching occurred around ±3 kOe can be explained that the applied perpendicular field is larger/smaller than the effective field of AFM coupling, which induces the switching of BML. The switching around 0 Oe arises from the switching of TML, which simultaneously induces the switching of BML again because of the strong AFM coupling. The configurations of magnetization in the TML and BML at each field stage are illustrated in insets of Fig. 1(e). The current induced magnetization switching is shown in Fig. 2. The measurement setup is similar to that of previously reported SOT induced switching [2,14]. Under an in-plane external field ($H_y$), we first applied a 1 ms current pulse ($I_p$) with gradually varying amplitudes, and then applied a 0.3 mA dc current after removal of $H_y$ to detect the magnetization state after each $I_p$. As shown in Fig. 2(a), the magnetization can be completely switched between two states and the switching orientation depends on the direction of external fields. For example, when $H_y$ = +1 kOe, the switching loop is clockwise, which becomes anticlockwise when $H_y$ = -1 kOe. The critical current density for showing switching behaviors is about $4.5 \times 10^7$ A/cm$^2$, which is comparable with that for switching a single ferromagnet [4,14] even though the total thickness of ferromagnets in the SAF is about three times larger than that of the single ferromagnet. These results indicate that the perpendicular SAFs can also be efficiently switched like a single ferromagnet by SOTs.

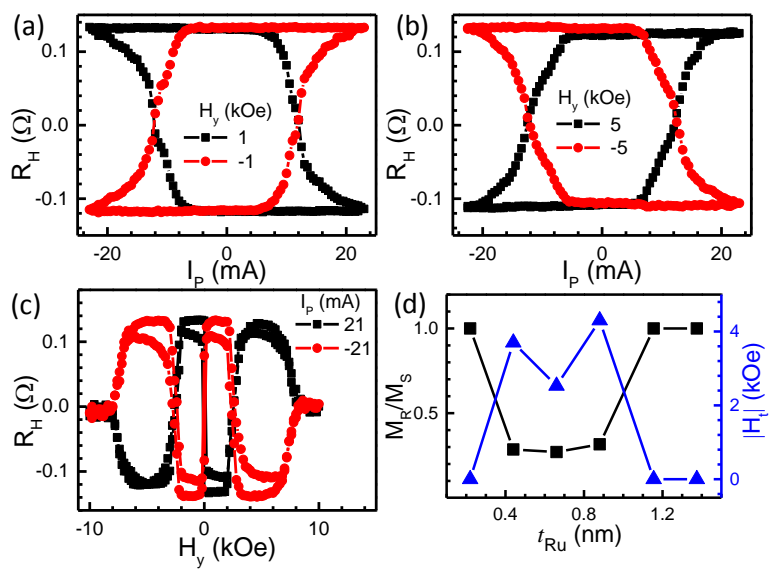

FIG. 2. The current driven magnetization switching under (a) ±1 kOe and (b) ±5 kOe in-plane magnetic fields. (c) The magnetization switching induced by ±21 mA current pulses as a function of in-plane field. (d) $M_R/M_S$ ratio (black) and $|H_t|$ (blue) as a function of $t_{Ru}$. $|H_t|$ = 0 indicates no AMS observed.

The surprising switching behavior occurs under a larger in-plane external field, as shown in Fig. 2(b). Now the switching loop changes the sign to be anticlockwise for +5 kOe and clockwise for -5 kOe, which is like the SOT switching with an opposite SHA sign. Figure 2(a) and (b) indicate that the magnetization can also be switched without any direction change of field or current. This unusual switching behavior has never been observed before and is quite different from all previous experimental results [2–4,8,10,12–17] and the macrospin based models [3,4,37–41]. As illustrated in Figs. 1(a) and (b) and widely verified in previous studies, the SOT switching orientation is only determined by the direction of applied in-plane field and the sign of SHA. Because only Pt with a positive SHA is involved in the sample, the switching orientation should keep the same if the direction of the applied magnetic field is not changed. To get a full switching phase of the sample, we measured the stable magnetization of the sample after current induced switching as a function of in-plane field, as shown in Fig. 2(c). In this measurement, the switching current pulse was maintained at ±21 mA while sweeping the in-plane field. It can be clearly seen that $R_H$ changes its sign at the transition fields ($H_t$) of ±2.5 kOe, in addition to the sign change around zero magnetic field that is expected from the conventional SOT switching. This magnetic field dependence of SOT switching further confirms the AMS in AFM coupled samples. Figure 2(d) shows the $t_{Ru}$ dependent SOT switching as well as the ratio between remanent magnetization ($M_R$) and saturation magnetization ($M_S$), in which $M_R/M_S \approx 0.3$ indicates AFM coupling and $H_t = 0$ Oe indicates no AMS observed. One can see that only the AFM coupled samples show AMS. Moreover, when $t_{Ru}$ approaches the values for FM coupling, the magnetization can only be partially switched at high field regions. As shown in Figs. 3(a) and (b), when $t_{Ru}$ = 0.44 nm and 0.88 nm, the magnetization after positive and negative switching currents only shows a slight change at high field regions. It should be noted that the partial switching is not due to the insufficient applied current. As shown in Figs. 3(a) and (b), even for larger applied currents of $I_p$ = ±25 mA, the switching loops keep the same as those of $I_p$ = ±21 mA.

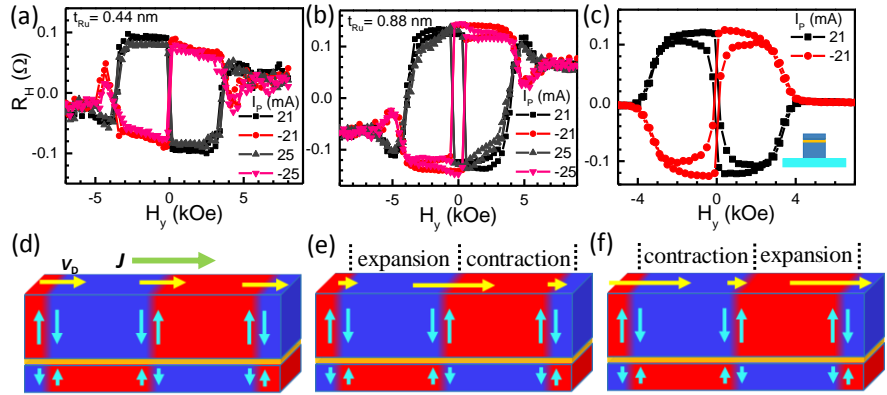

FIG. 3. $t_{Ru}$ dependent current induced magnetization switching with (a) $t_{Ru}$ = 0.44 nm and (b) 0.88 nm. (c) The current induced magnetization switching in a control sample with a thicker BML. Inset of (c) shows the relative thickness of BML and TML. (d-f) Illustrations of SOT switching based on DW motion. (d) All DWs move with the same velocity, and domains keep the same shape during the current driven DW motion. (e, f) The different velocities of ↑↓$_{BML}$ and ↓↑$_{BML}$ DWs induce the expansion or contraction of opposite domains during the current driven DW motion. Yellow arrows represent DW velocities.

To understand the mechanism of this unusual switching behavior, the SOT switching was also studied in a control sample of Pt (4nm)/Co (0.8nm)/Ru (0.66nm)/Co (0.2nm)/Pt (1nm)/Co (0.4nm)/Ru (0.6nm)/SiO$_2$ (10nm) with a thicker BML. Even with the same SOT and $t_{Ru}$, the AMS is missing in the control sample as shown in Fig. 3(c) and the switching behaviors consist with previous reports [2,4]. All of these results indicate that two conditions must be met to exhibit AMS: (1) AFM coupling and (2) the BML is thinner than the TML. The observed AMS cannot be explained by the macrospin model, in which the SOT switching of a strongly coupled SAF is the same as that of a single FM with an effective magnetization **M**$_{eff}$ (see Supplemental Material). In addition, the partial switching in the high field regions as shown in Figs. 3(a) and (b) also sheds light on the violation of the macrospin model, because the partial switching should gradually evolve to a full switching by increasing the applied current.

## IV. Discussion

Recently, it has been experimentally observed that an in-plane field can result in asymmetric domain expansion in HM/FM structures [12,23–25,42–44]. Here we propose a switching model to explain the AMS directly arising from the asymmetric domain expansion. The simple picture of this model (one-dimensional) is given in Figs. 3(d-f). First, without external fields, the applied current induces a demagnetized state and drives all DWs to move with the same velocity ($V_D$). The demagnetized state is induced by the combination of all spin torques and thermal effects which has been proven to destabilize a uniform magnetization [45–48] and finally leads to a demagnetized state with equal spin-up and spin-down domains. The applied in-plane field may also assist with the formation of the demagnetized state but mainly modulates the DW motion, inducing the asymmetric expansion/contraction of a domain as shown below. In this case, the domain shape and area keep the same during DW motion and no favored magnetized direction is formed (Fig. 3(d)). Second, if an applied field can separately modulate the velocities of ↑↓$_{BML}$ and ↓↑$_{BML}$ DWs by increasing the $V_D$ of one type of DW and decreasing that of another type of DW, the domains will expand or contract during the current driven domain motion, as shown in Figs. 3(e) and (f). Now the magnetization will favor either spin-up or spin-down states, depending on the relative velocity between two DWs, $V_{RD}$. Third, to reach a full magnetization switching, $V_{RD}$ has to be large enough to collapse those contracted domains and any possible nucleated reversal domains within the expanded domains. This is because a reversal domain will nucleate again within the expanded domain to keep the demagnetized state if the expanded domain is larger than a critical value. Above the critical value, a domain can still be thought as a uniform magnetization with higher magnetic energy that is unstable under the large current. As demonstrated before, the separate control of $V_D$ for two types of DWs can be realized in HM/FM bilayers and $V_{RD}$ depends on the strength of applied in-plane fields [19,20], therefore, a large enough field is necessary to realize a full SOT switching [2–4,27] in these structures.

According to this model, the magnetization switching orientation is only determined by the sign of $V_{RD}$, regardless of the sign of the SHA and the initial domain nucleation directions. This is because, no matter the SHA is positive or negative, SOTs will lead to the same demagnetized state (first condition) and the second and third conditions are only determined by $V_{RD}$. This is a distinct difference between this model and previous models [3,4,37–41]. Although previous models also suggested that the switching process can be incoherent, the sign of the SHA still determines the nucleation direction of the first domain according to the macrospin switching model, and thus decides the final switching orientations [12,23–25] because the subsequent switching process is based on the expansion of the initially nucleated domains. One can see that the role of in-plane field in this model are also very different from previous models where the external field was used to stabilize magnetization or break symmetry [2–4,16].

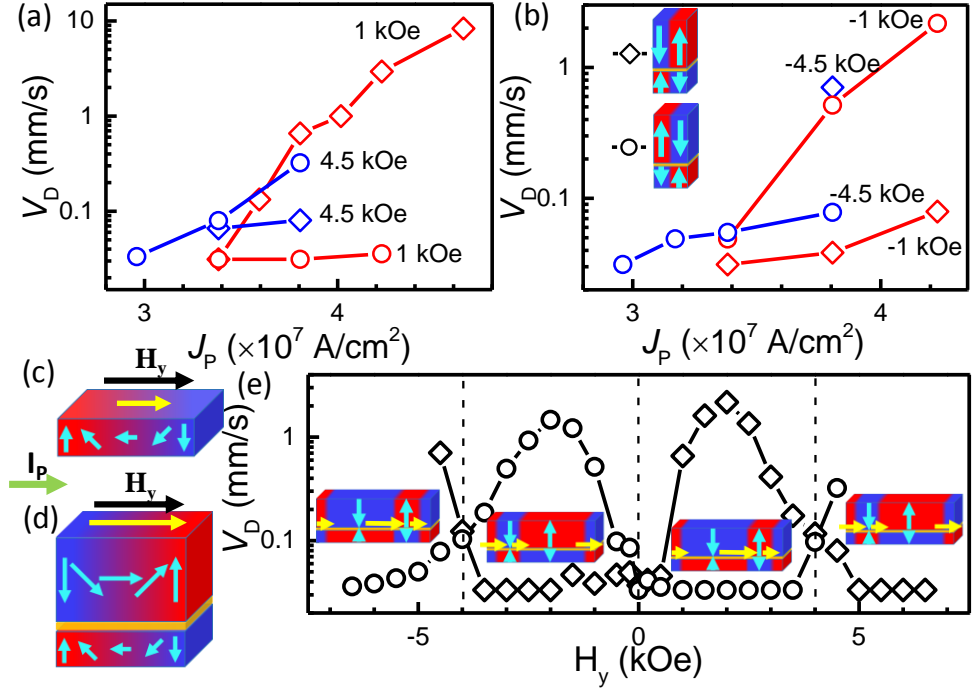

FIG. 4. The current driven DW motion as a function of current density under (a) positive and (b) negative in-plane fields. Red and blue represent ±1 kOe and ±4.5 kOe external fields, respectively. (c, d) The schematic of in-plane field-modulated DW motion for (c) a single magnetic layer and (d) AFM coupled bilayer. (e) $V_D$ as a function of in-plane field driving by the pulses with the current density of $3.97\times10^7$ A/cm$^2$. The inserted illustrations illustrate the domain expansion/contraction at each in-plane field region, consistent with the four SOT switching regions shown in Fig. 2(c). For all figures, diamonds and circles represent ↑↓$_{BML}$ and ↓↑$_{BML}$ DWs, respectively. For clarification, the error bars are omitted (See Supplemental Material for the determination of error).

To verify this model, we measured $V_D$ in our sample. Figures 4(a) and (b) present $V_D$ as a function of applied current when $H_y = \pm 1$ kOe and $\pm 4.5$ kOe, which show that $V_D$ increases with applied current density for both types of DWs. When $H_y = +1$ kOe, $V_{RD}$ between ↑↓$_{BML}$ and ↓↑$_{BML}$ DWs is positive, and thus the spin-up state of BML is favored for $+I_P$ according to the illustrations of Fig. 3(e). This is completely consistent with the switching orientation shown in Figs. 2(a) and (c). In contrary, when $H_y = +4.5$ kOe, $V_{RD}$ becomes negative, and consequently, the spin-down state of BML is favored for $+I_P$, also consistent with the switching in Fig. 2(c). When either $H_y$ or $I_P$ changes sign, the switching orientations in Fig. 2(c) can also be understood through the sign change of $V_{RD}$. Figure 4(e) gives the measured $V_D$ for two types of DWs at different in-plane fields. The four regions with positive or negative $V_{RD}$ are consistent with the four switching regions of Fig. 2(c), confirming our explanations. It should be noted that the $V_{RD}$ changes sign around ±4 kOe, which is larger than $H_t$ shown in Fig. 2(c). This is because the DW velocity was measured at $I_p = \pm9$ mA and the resultant thermal effect is much smaller than that of $I_p = \pm21$ mA used for SOT switching, resulting in a larger $H_t$ (see Supplemental Material).

As shown in Fig. 4(e), the modulation effects of $H_y$ on $V_D$ are quite different from those in HM/FM structures [19,20]. In HM/FM structures, $V_{RD}$ changes sign only once around $H_y = 0$ Oe [19,20]. Therefore, according to our model, the switching orientation only reverses once around 0 field, consistent with the experimental results [2–4]. In our samples, $V_{RD}$ changes sign three times, corresponding to the four contrasting switching regions as shown in Fig. 2(c). Furthermore, in HM/FM structures, the switching orientation due to $V_{RD}$ is the same as that predicted by the macrospin model [4]. While in our samples, $V_{RD}$ changes sign even for the same field direction, and the switching orientation contradicts the macrospin model, clearly clarifying the AMS as arising from field modulated $V_{RD}$. The partial switching shown in Figs. 3(a) and (b) may be due to a small $V_{RD}$, which is not large enough to collapse all nucleated reversal domains within the expanded domains.

The unique field modulation effects in our structures can be attributed to the strong AFM coupling. It is shown that the field modulation of $V_D$ is determined by the parallel/antiparallel configuration between $H_y$ and the internal magnetization of DWs [19,20]. As shown in Fig. 4(c), in a single FM (also the same for two FM coupled magnetic layers, or two AFM coupled layers but with a thicker BML, see Supplemental Material), the transition between the parallel and antiparallel configurations only occurs once when $H_y \approx -H_{DMI}$, where $H_{DMI}$ is the DMI effective field ($H_{DMI}$) [19,20]. However, for a SAF with a thicker TML (Fig. 4(d)), the **M**$_{eff}$ is determined by the thicker TML layer. Correspondingly, the transition between the parallel and antiparallel configurations occurs twice when $H_y \approx -(H_{DMI} + H_{exc})$ (see Supplemental Material), resulting in the unique $H_y$ modulated $V_{RD}$. Here, $H_{exc}$ is the effective AFM coupling field in BML, and we ignore the SOTs and $H_{DMI}$ in TML.

The understanding of AMS based on the domain nucleation and DW motion indicates that the DW energy dominates the magnetization dynamics in such structures. In multilayers with PMA, magnetic domains or skyrmions can be formed with very high density (indicating high DW energy) during SOT induced magnetization dynamics because of the narrow size of DWs [49], and thus the involvement of DW energy is required to improve previous switching models [3,4,37–41] to explain AMS. As demonstrated here, the DW energy may dominate the magnetization dynamics compared with other contributions (such as the sign of SOT). The recently reported memristive behaviors in antiferromagnet-ferromagnet bilayers [10] are probably also due to a small $V_{RD}$ like the partial switching in Figs. 3(a) and (b). Because of the narrow DWs (typical several nm [49]), the reversal mechanism demonstrated here will still dominate the switching process for the feature size of tens nm in future SOT-MRAMs.

**V. Summary**

In summary, we have demonstrated the SOT switching of SAFs, which shows an anomalous switching behavior compared with a single ferromagnet. These results offer new possibilities to explore SOT related magnetization dynamics in magnetically coupled multilayers and also clarify the SOT switching mechanism. Although plenty of physical phenomena such as the Rashba effects, SHEs, and DMI effects, have been observed in the HM/FM interfaces after the discovery of SOT switching, how these effects directly contribute to the SOT switching is not clear. In the macrospin model, only the damping-like torques from SHEs was included. The demonstrated new switching model based on the chiral DW motion clearly indicates that all those interfacial effects contribute to the SOT switching indirectly by driving and modulating DW motion. The field-like torques that can also originate from the Rashba effects and dramatically modulate the DW motion [50] may also determine the SOT switching. This switching model also highlights the essential role of DMI effects, which are the origin of DW chirality [19,20], in the SOT switching.

In application, the demonstrated SAF switching will benefit the high-density SOT-MRAMs by addressing the emerging challenges in nanosized MTJs with high thermal stability and efficient switching. In addition, it has been shown theoretically and experimentally that the interlayer coupling can be changed between AFM and FM states by voltage [34–36]. Together with the sign control of SOT switching demonstrated here and a very large DW velocity realized as the magnetization of BML and TML approach each other [51], a voltage tunable high-speed, low-energy manipulation of magnetization could possibly be realized in the HM/SAF structures.

**Acknowledgement**

This work was supported in part by C-SPIN, one of six centers of STARnet, a Semiconductor Research Corporation program, sponsored by MARCO and DARPA; and by the National Science Foundation through ECCS-1310338.

[48] J. Torrejon, F. Garcia-Sanchez, T. Taniguchi, J. Sinha, S. Mitani, J.-V. Kim, and M. Hayashi, Phys. Rev. B **91**, 214434 (2015).

[49] A. Thiaville, S. Rohart, É. Jué, V. Cros, and A. Fert, EPL (Europhysics Lett. **100**, 57002 (2012).

[50] I. M. Miron, T. Moore, H. Szambolics, L. D. Buda-Prejbeanu, S. Auffret, B. Rodmacq, S. Pizzini, J. Vogel, M. Bonfim, A. Schuhl, and G. Gaudin, Nat. Mater. **10**, 419 (2011).

[51] S.-H. Yang, K.-S. Ryu, and S. Parkin, Nat. Nanotechnol. **10**, 221 (2015).

.

Supplemental Material

# Spin-orbit torque switching of synthetic antiferromagnets

Chong Bi[1], Hamid Almasi[1], Kyle Price[1], T. Newhouse-Illige[1], M. Xu[1],

Shane R. Allen[2], Xin Fan[2] and Weigang Wang[1*]

[1]Department of Physics, University of Arizona, Tucson, Arizona 85721, USA

[2]Department of Physics and Astronomy, University of Denver, Colorado 80208, USA

* wgwang@email.arizona.edu

1. **DW motion measurements**
2. **Macrospin model of SOT switching in synthetic ferromagnets/antiferromagnets**
3. **Angle dependent spin-flop fields**
4. **Current induced DW motion in SAFs with a thicker TML**
5. **Thermal effects on AFM coupling**

## 1. DW motion measurements

To measure the DW motion, we first applied a nucleation pulse to create a domain in our samples. As shown in Fig. S1(a), if the domain was successfully created, the switching field of $R_H$ reduced about 30 Oe. Fig. S1(b) shows the change of $R_H$ as a function of the number of driving current pulses with/without domain nucleation. One can see that the increase of $R_H$ with the number of injected 10 mA current pulses is due to the DW motion because it was only observed after a domain was created. If we apply a large driving current (>15 mA), the $R_H$ also changes as a function of pulse number regardless of the domain creation (not shown), indicating that the change of $R_H$ in this case is due to the current induced domain nucleation rather than DW motion. We found that the threshold of driving current for domain nucleation strongly depends on the external fields, even on the in-plane fields. In the subsequent measurements, we needed to avoid the current induced domain nucleation, which was determined experimentally by measuring if $R_H$ changed without domain creation. Fig. S1(c) shows the DW motion under different pulses. For the pulses less than 8 mA even with a longer length, the DW cannot be moved. The current threshold for DW motion also depends on the external fields. Fig. S1(d) shows the DW motion under pulses with the same amplitude but with different lengths. One can find that the time for DW motion between the nucleation line and the voltage bars varies a little bit with the length of the applied current pulse. Therefore, in addition to reducing the driving current density to make sure that the time scale was within 1ms to 10s as mentioned above, we also tried to make the entire DW motion within one pulse to increase the accuracy of our measurements as much as possible. The measurement processes are as follows:

1) Applying a large perpendicular field to initialize the magnetization to a uniform up/down state.
2) A 120 mA nucleation current pulse was applied to create a domain. During the domain creation, a perpendicular assisting field was also applied as mentioned above.
3) Applying a 10 s current pulse to driven DW motion, and then detecting $R_H$ to determine if the DW reached the voltage bars.
4) If $R_H$ changed to a full reversal state after the 10 s current pulse, which indicates that 10 s is enough for DW motion between the nucleation line and the voltage bars, we repeated 1) - 3) but reduced the length of pulse to half of the previous value (for example, 5 s for the second time). Similarly, we increased the length of injected pulse if $R_H$ was not changed.
5) Repeated 1) – 4) until the time error was within 20% (for example, finally, we found $R_H$ changed after a 100 ms pulse but did not change after an 80 ms pulse, and then we chose 90 ms as the time for calculating DW velocity) or less than 1 ms.
6) For each condition, we measured ten times and then extracted the average value and error as the DW velocity.

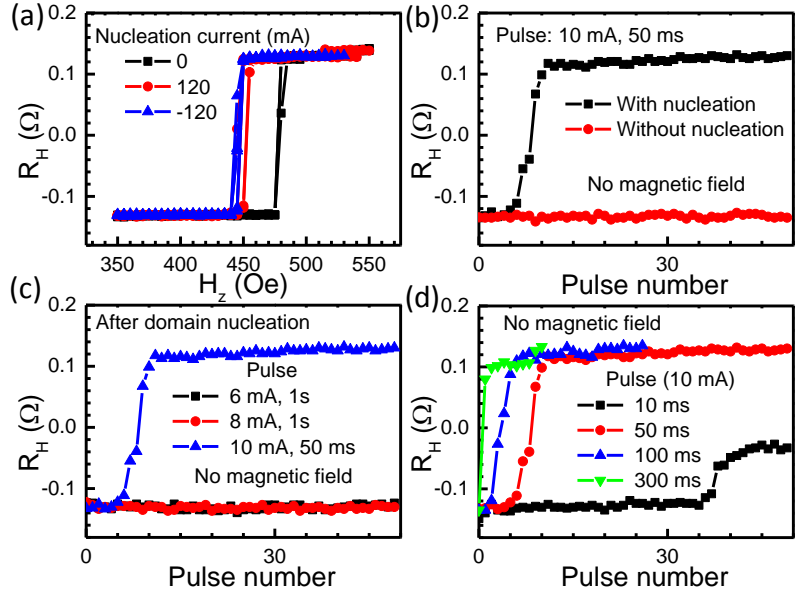

FIG. S1. (a) Perpendicular field driven magnetization switching from a uniform down state with/without domain creation in the Hall bar. (b) The evolution of magnetization state in the cross of voltage bars after the 10 mA, 50 ms current pulses with/without domain creation. (c) $R_H$ as a function of the pulse number of driving current after the domain creation. The applied current pulses are 6 mA, 1 s (black), 8 mA, 1 s (Red), and 10 mA, 50 ms (blue). (d) $R_H$ as a function of the number of 10 mA current pulses with different lengths.

## 2. Macrospin model of SOT switching in synthetic ferromagnets/antiferromagnets

Generally, the magnetization dynamics of a synthetic ferromagnet/antiferromagnet with a strong interlayer coupling can be considered the same as that of a single ferromagnet with the effective magnetization of $\mathbf{M}_{\text{eff}} = \mathbf{M}_{\text{B}} + \mathbf{M}_{\text{T}}$, where $\mathbf{M}_{\text{B}}$ and $\mathbf{M}_{\text{T}}$ are the magnetization of BML and TML, respectively. If the antiferromagnet is thought of as a single ferromagnet with $\mathbf{M}_{\text{eff}}$ in macrospin model, the SOT switching will be the same as that in a single ferromagnet, which cannot explain the observed AMS. However, even if we consider the two coupled ferromagnetic layers separately, the macrospin model still gives the same SOT switching as a single ferromagnet with $\mathbf{M}_{\text{eff}}$.

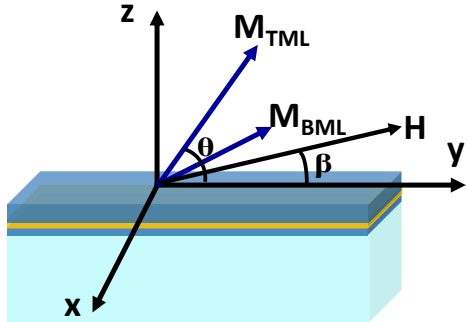

FIG. S2. Illustration of coordinate system used in macrospin simulations.

As shown in Fig. S2, we consider two coupled ferromagnets adjacent to a heavy metal with a positive SHA and ignore the SOTs in TML. The total spin torques for BML and TML can be written as

$$\tau_{BML} = \tau_{SHE} + \tau_{ext}^{BML} + \tau_{an}^{BML} + \tau_{exc}^{BML}$$
$$= H_{SHE}(\mathbf{M_B} \times \mathbf{x} \times \frac{\mathbf{M_B}}{M_B}) - \mathbf{M_B} \times \mathbf{H} - \mathbf{M_B} \times \mathbf{H_{KB}} - \mathbf{M_B} \times \mathbf{H_{exc}^B},$$

and

$$\tau_{TML} = \tau_{ext}^{TML} + \tau_{an}^{TML} + \tau_{exc}^{TML}$$
$$= -\mathbf{M_T} \times \mathbf{H} - \mathbf{M_T} \times \mathbf{H_{KT}} - \mathbf{M_T} \times \mathbf{H_{exc}^T},$$

respectively, where $H_{SHE}$ is the effective field arising from SOTs, $\mathbf{H_{KB}}$ and $\mathbf{H_{KT}}$ are the perpendicular anisotropy fields of BML and TML, respectively. The interlayer coupling is described through the exchange fields of $\mathbf{H_{exc}^B}$ and $\mathbf{H_{exc}^T}$ in the BML and TML, respectively, which are derived from $\mathbf{H_{exc}^{B/T}} = -\frac{\partial E^{exc}}{\partial \mathbf{M_{B/T}}}$, where $E^{exc} = -2J^{exc}\frac{\mathbf{M_B} \cdot \mathbf{M_T}}{M_B M_T}$ is the exchange energy. $J^{exc}$ is positive and negative for FM and AFM coupling, respectively. For a steady state, the simulation results are shown in Fig. S3-S5, where the external field is applied along y direction, $H_{KB} = H_{KT}$, and $H_{exc}^B = -3H_{KB}$ and $3H_{KB}$ for AFM coupling and FM coupling, respectively. Figure S3 shows the SOT switching for an AFM coupled system with a thicker TML ($M_T/M_B = 2$). The SOT switching orientations of BML and TML are opposite in this system, and the switching orientation of $\mathbf{M_{eff}}$ (Figs. S3(c) and (f)) is the same as that of a single ferromagnet [1]. One can see that the switching orientation does not depend on the strength of $H_y$. We also simulated the switching for a weaker $H_{exc}^B = -0.5H_{KB}$ or stronger $H_{exc}^B = -10H_{KB}$ coupling system, and no $H_y$ dependence of the switching orientation was observed. Figure S4 shows the SOT switching of AFM coupled systems under $H_y = \pm 0.5H_{KB}$ with different relative thickness of the BML and the TML. The switching of BML changes orientation when BML is thicker than TML, but the switching orientation of $\mathbf{M_{eff}}$ keeps the same as that of a single ferromagnet. Figure S5 shows the SOT switching of a FM coupled system with $M_T/M_B = 2$, in which the switching orientations of both BML and TML are the same as those in a single ferromagnet. All of these results support that a synthetic ferromagnet/antiferromagnet with a strong interlayer coupling can be thought as a single ferromagnet with the effective magnetization of $\mathbf{M_{eff}}$ in the SOT driven switching, which cannot explain our observed AMS.

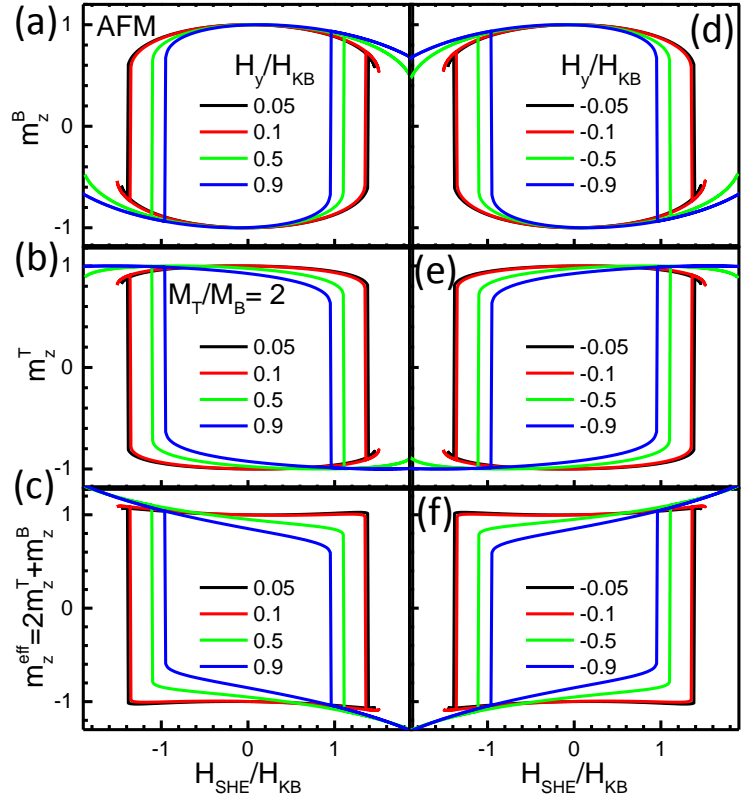

FIG. S3. The simulated SOT switching of an AFM coupled system with a thicker TML by the macrospin model. The top panels show the SOT switching of BML under (a) +$H_y$ and (d) -$H_y$. The medium panels show the SOT switching of TML under (b) +$H_y$ and (e) -$H_y$. The bottom panels show the switching of effective magnetization $m_z^{eff} = 2m_z^T + m_z^B$ under (c) +$H_y$ and (f) -$H_y$, which is the same as that of a single ferromagnet.

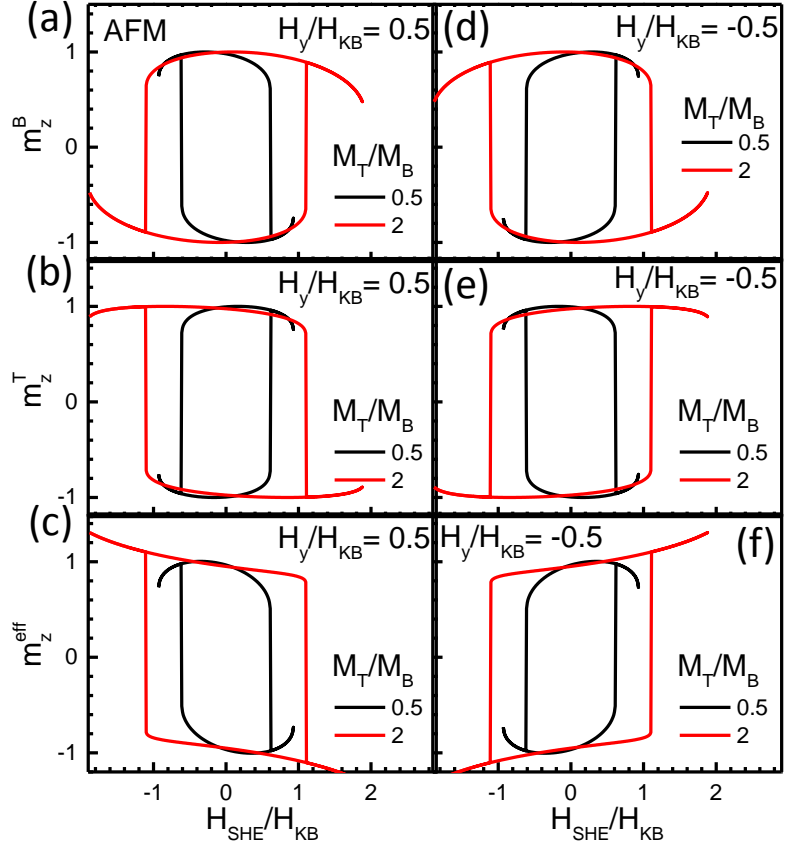

FIG. S4. The simulated SOT switching of an AFM coupled system with different relative thicknesses between BML and TML by the macrospin model. The top panels show the SOT switching of BML under (a) $H_y = 0.5 H_{KB}$ and (d) $H_y = -0.5 H_{KB}$. The medium panels show the SOT switching of TML under (b) $H_y = 0.5 H_{KB}$ and (e) $H_y = -0.5 H_{KB}$. The bottom panels show the switching of effective magnetization under (c) $H_y = 0.5 H_{KB}$ and (f) $H_y = -0.5 H_{KB}$. For $M_T/M_B = 2$, $m_z^{eff} = 2m_z^T + m_z^B$; For $M_T/M_B = 0.5$, $m_z^{eff} = m_z^T + 2m_z^B$.

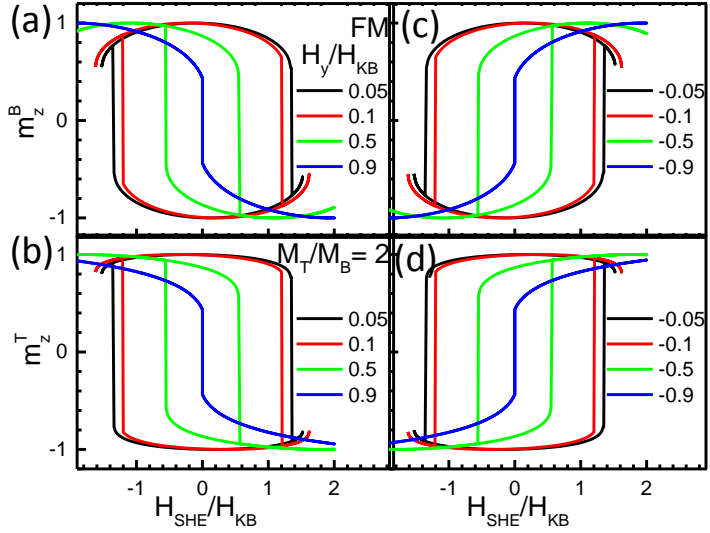

FIG. S5. The simulated SOT switching of a FM coupled system by the macrospin model. The top panels show the SOT switching of BML under (a) +$H_y$ and (c) -$H_y$. The bottom panels show the SOT switching of TML under (b) +$H_y$ and (d) -$H_y$. The switching orientations of both layers are consistent with those of a single ferromagnet.

### 3. Angle dependent spin-flop fields

Although the possible spin-flop transition under external fields has been described in the above macrospin model, we measured the angle dependent spin-flop fields ($H_{SF}$, defined as shown in Fig. S6(a)) experimentally as shown in Fig. S6. Figure S6(a) shows $R_H$ as a function of external field at different angles. One can see that $H_{SF}$ keeps almost constant when $45° < \beta < 90°$ and dramatically increases when $\beta$ approaches $0°$. Figure S6(b) gives $H_{SF}$ as a function of $\beta$ and the red solid lines are simulated results through $H_{SF} = \dfrac{\left.H_{SF}\right|_{\beta=90°}}{\sin(\beta)}$, where $\left.H_{SF}\right|_{\beta=90°} = \pm 3\,\text{kOe}$. All of these results indicate that the spin-flop transition occurs when the perpendicular component of applied field is larger than the effective AFM coupling field. In our current-induced SOT switching measurements, the external field is applied around $\beta = 0°$, at which the $H_{SF}$ is much larger than our applied field (0 – 10 kOe, as shown in Fig. 2(c)) and thus there is no spin-flop transition during our SOT switching measurements.

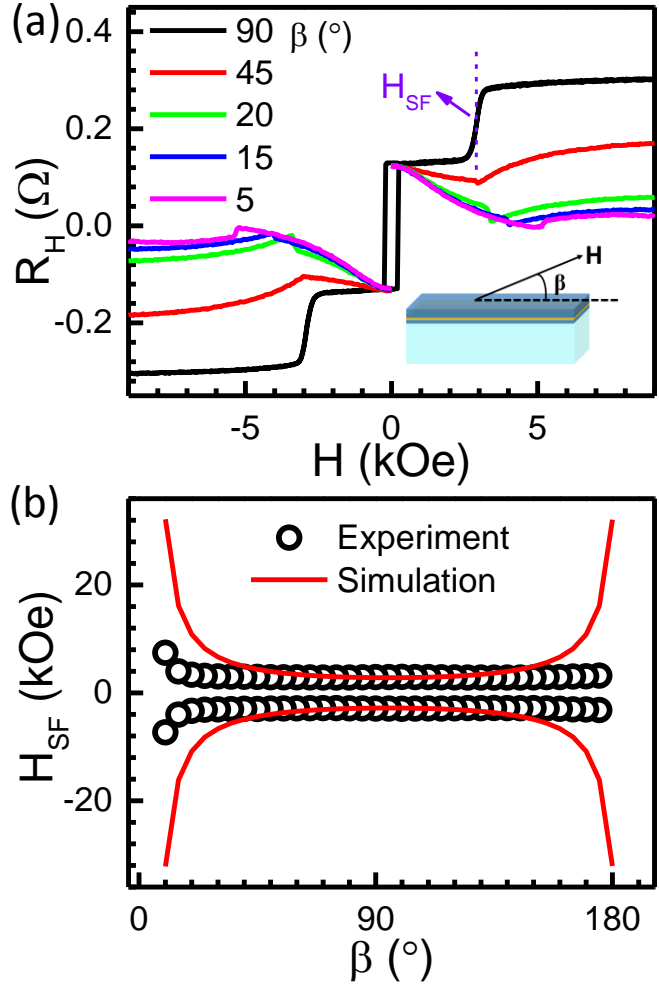

FIG. S6. (a) $R_H$ as a function of external field at different angles. The inset shows the definition of β. (b) The measured spin-flop field (black circles) as a function of β. The red solid lines are simulated angle dependent $H_{SF}$.

## 4. Current induced DW motion in SAFs with a thicker TML

In a ferromagnet adjacent to heavy metals, it has been shown that the current induced DW motion can be modulated by an applied in-plane field [2,3]. As shown in Fig. S7(a, b), without in-plane field, both the ↑↓ and ↓↑ DWs (Left-hand chirality) move with the same velocity driven by an in-plane current. Because of SOTs from the heavy metals and the interfacial Dzyaloshinskii–Moriya interaction (DMI), the DWs move along with the current direction, and the internal magnetization of the DWs is opposite for ↑↓ and ↓↑ DWs, along the -y and +y direction, respectively. When $H_y$ is applied, the DW motion will be promoted or suppressed depending on the parallel or antiparallel configuration between the internal magnetization and $H_y$. Previous results show that, when $H_y$ is antiparallel to the internal magnetization, the velocity of DWs will decrease and even change sign when $|H_y| > |H_{DMI}|$ [2,3], where $H_{DMI}$ is the interfacial DMI field. For example, for ↑↓ DWs, the internal magnetization and $H_{DMI}$ are along with -y

direction, thus a positive $H_y$ will decrease the velocity of ↑↓ DWs and even changes its sign when $H_y > -H_{DMI}$, as confirmed by the experimental results (Fig. S7(e)) [2,3]. For a FM coupled bilayer between two ferromagnets, or an AFM coupled bilayer but with a thicker BML, $H_{DMI}$ and the effective internal magnetization ($M_{eff}$), and thus the field modulated DW motion, are the same as those in a ferromagnet, because both $H_{DMI}$ and $M_{eff}$ are determined by the BML adjacent to the heavy metals.

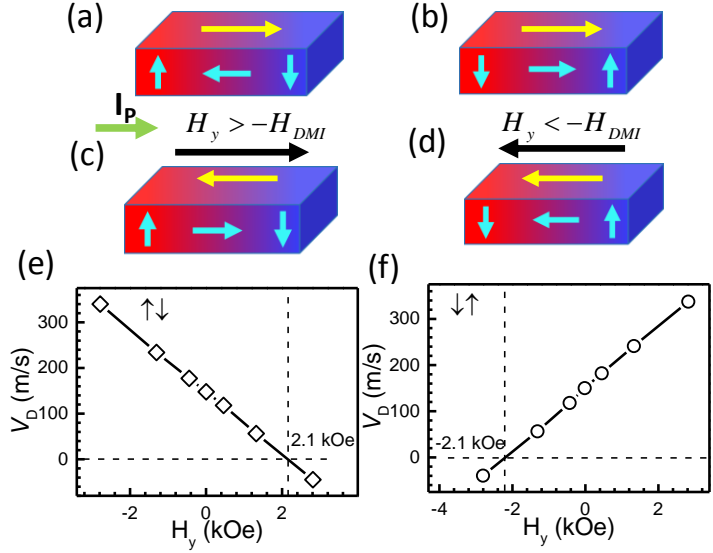

FIG. S7. Illustrations of the in-plane field modulated DW motion in HM/FM structures. (a, b) Without $H_y$, ↑↓ and ↓↑ DWs have the same velocity under an in-plane driving current. (c, d) When applying an antiparallel in-plane field with the internal magnetization of DWs, the velocity of DWs decreases and even changes sign if $|H_y| > |H_{DMI}|$ due to the change of DW chirality from left-hand to right-hand. (e, f) The experimental data [2] of DW velocity (under a positive driving current) as a function of in-plane field for (e) ↑↓ and ↓↑ (f) DWs, which is consistent with the illustrations of (c) and (d), respectively.

However, for an AFM coupled system but with a thicker TML (satisfying the two conditions for showing AMS, as mentioned in text), the situation will be different. The pronounced feature of this system is that the $M_{eff}$ of the two coupled DWs is determined by the TML, which indicates that the internal magnetization reversal of the DWs in the BML is not only determined by $H_y$, but also strongly relates to the AFM coupling field ($H_{exc}$). As illustrated in Figs. S8(a, c), when $|H_y| > |H_{DMI} + H_{exc}|$, $H_y$ can cancel $H_{DMI}$ and $H_{exc}$, resulting in the parallel configuration of internal magnetization between the BML and the TML. Now the configurations of the internal magnetization of the DWs in the BML and TML have four states, as shown in insets of Figs. S8(b, d). The velocity as a function of in-plane field shown in Figs. S8(b, d) is from Fig. 4(e). The configuration transitions between parallel and antiparallel around -3.5 kOe and 5 kOe (Fig. S8(b)) are caused by a larger/smaller $H_y$ than $H_{DMI} + H_{exc}$, and the transition between two antiparallel configurations around 0 Oe is due to the $H_y$ induced effective internal magnetization switching of the two coupled DWs in the BML and TML. One can see that the corresponding internal magnetization of DWs in the BML is switched three times in entire $H_y$ range. If we only consider the SOT driven DW motion in the BML, each reversal of the internal magnetization of DWs in the BML will lead to a dramatic change of the DW velocity, which is completely consistent with our experimental results as shown in Fig. S8(b, d). Because the dramatic changes of velocity around -3.5 kOe and 5 kOe (Fig. S8(b)) correspond to $H_{DMI} + H_{exc}$, we can evaluate that $H_{DMI}$ = -0.8 kOe and $|H_{exc}|$ = 4.3 kOe for ↑↓$_{BML}$ DWs. Similarly, $H_{DMI}$ = 1.0 kOe and $|H_{exc}|$ = 4.5 kOe for ↓↑$_{BML}$ DWs from Fig. S8(d). The evaluated $H_{exc}$ value approaches that measured by AHE and VSM as shown in Fig. 1(e), and the $H_{DMI}$ values are also reasonable compared with previous reports [2,3].

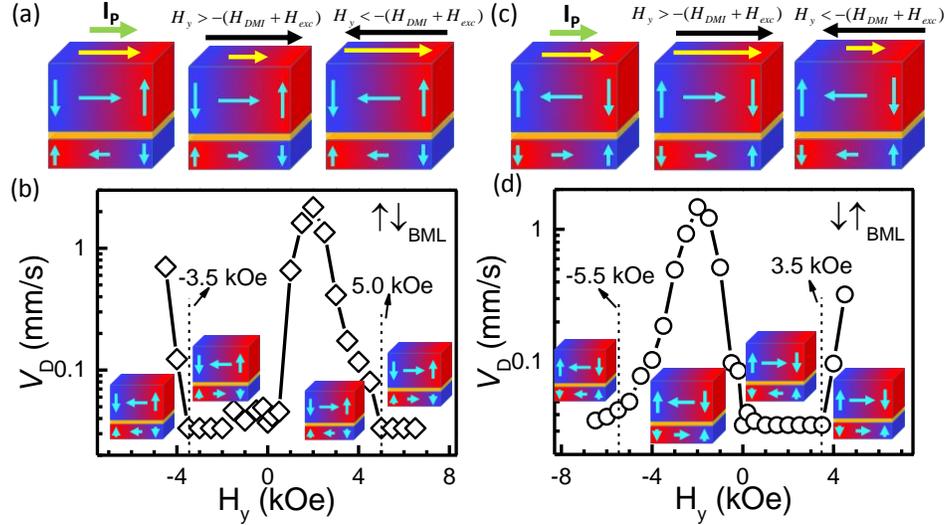

FIG. S8. Illustrations of the in-plane field modulated DW motion in an AFM coupled bilayer with a thicker TML. (a, c) Illustrations of parallel/antiparallel configurations of internal magnetization of DWs in BML and TML, in which the antiparallel configuration occurs when $|H_y| > |H_{DMI} + H_{exc}|$. (b, d) The experimental results of velocity as a function of in-plane field (from Fig. 4(e)) for (b) ↑↓$_{BML}$ and (d) ↓↑$_{BML}$ DWs.

Figure S9 shows $V_{RD}$ as a function of $H_y$ under a positive driving current for HM/FM (Fig. S9(a)) and HM/SAF structures with a thicker TML (Fig. S9(b)). As mentioned in our model, the $V_{RD}$ decides the magnetization switching direction. For HM/FM structures, the sign of $V_{RD}$ keeps the same for the same field direction, while for HM/SAF structures, $V_{RD}$ also changes its sign around $|H_{DMI} + H_{exc}|$. The signs of $V_{RD}$ in both systems are consistent with the corresponding SOT switching orientations under different $H_y$ as expected in our model.

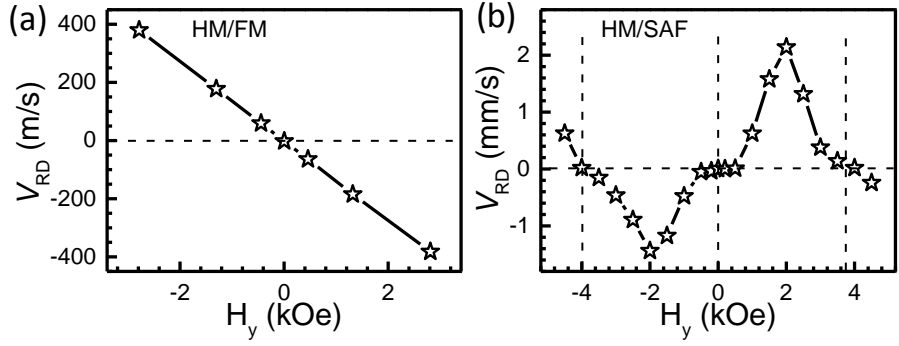

FIG. S9. The experimental relative velocity ($V_{RD}$) between ↑↓ and ↓↑ DWs (in BML for SAF) as a function of $H_y$ for (a) HM/FM and (b) HM/SAF structures. The driving current is positive. The data for HM/FM and HM/SAF structures is from Fig. S7(e, f) and Fig. S8(b, d), respectively. For the HM/SAF structures, $V_{RD}$ also changes sign around $|H_{DMI} + H_{exc}|$, consistent with the observed four switching regions (as shown in Fig. 2(c)) in this system.

### 5. Thermal effects on AFM coupling

To evaluate the Joule heating on the AFM coupling field, we measured perpendicular $R_H$ loops under different applied dc currents (I), as shown in Fig. S10(a). With increasing the applied current, the spin-flop transition gradually shifts toward 0 Oe. On the other hand, the shift of the spin-flop transition does not depend on the current direction (see ±15 mA curves in Fig. S10(a)), indicating a pure thermal effect. We choose the field at which the magnetization of the BML and TML become fully parallel as $H_{exc}$, as defined in Fig. S10(a). Figure S10(b) shows that $H_{exc}$ has a linear relation with $I^2$, as excepted by thermal effects. These results demonstrated that the large applied current will reduce the $H_{exc}$ due to Joule heating.

In SOT switching measurements, the applied current is 21 mA (Fig. 2(c)), which is much larger than the applied current for measuring DW motion (9 mA in Fig. 4(e)). According to our switching model, the transition field $H_t$ between two switching orientations in Fig. 2(c) is the same as the transition field for the sign change of $V_{RD}$, which corresponds to $H_{DMI} + H_{exc}$. Therefore, a larger current will result in a smaller $H_t$ due to the thermal effects, which is consistent with our observations in Fig. 2(c) and Fig. 4(e).

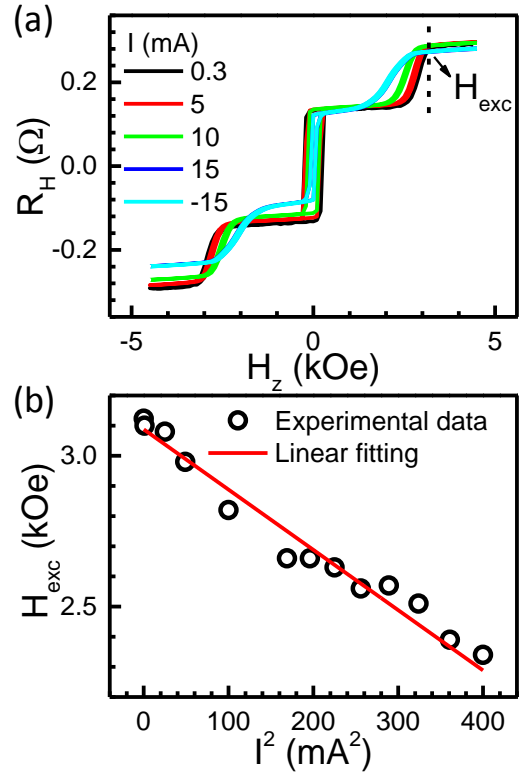

FIG. S10. (a) Perpendicular $R_H$ loops measured at different currents. (b) The measured $H_{exc}$ as a function of $I^2$. The red line is linear fitting.